\documentclass[a4paper,12pt]{article}
\usepackage{jheppub}

\usepackage{etoolbox}
    \makeatletter
    \patchcmd{\maketitle}{\@fpheader}{}{}{}
    \makeatother

\usepackage{amsmath}
\usepackage{amsthm}
\usepackage{amsfonts}
\usepackage{amssymb}
\usepackage{hyperref}
\usepackage{textcomp}
\usepackage{graphicx}
\usepackage{bm}
\usepackage{color}
\usepackage{verbatim}

\usepackage{graphicx,color}
\usepackage{epstopdf}
\usepackage{epsfig}

\usepackage{amsfonts,amsmath,amssymb}

\newcommand{\p}{\partial}

\newcommand{\rar}{\rightarrow}

\newcommand{\lvec}{\overrightarrow}

\begin{document}

\title{\textbf{Numerical Solution \\ of the Boundary Value Problems \\ for Partial Differential Equations. }\\ Crash course for holographer}
\date{}

\author{Alexander Krikun}
\affiliation{ Instituut-Lorentz, Universiteit Leiden, Delta-ITP \\ 
 \textit{P.O. Box 9506, 2300 RA Leiden, The Netherlands}
} 

\abstract{These are the notes for a series of Numerical Study group meetings, held in Lorentz institute in the fall of 2017. The aim of the notes is to provide a non-specialist with the minimal knowledge in numerical methods used in BVP for PDEs, necessary to solve the problems typically arising in applications of holography to condensed matter systems. A graduate level knowledge of Linear Algebra and theory of Differential Equations is assumed. Special attention is payed to the treatment of the boundary conditions of general form. The notes focus on the practical aspects of the implementation leaving aside the theory behind the methods in use. A few simple problems to test the acquired knowledge are included.}

\maketitle

\newpage

\section{Introduction}
The applications of AdS/CFT to condensed matter physics have passed the stage of the ``proof of concept'' and as the problems under the focus become more involved, the more sophisticated numerical machinery is needed to track them. This sets a considerable obstacle to the community of holographers who, being trained as theoretical physicists, often lack the necessary numerical skills. The goal of these notes is to provide a detailed tutorial, to those willing to learn how to use numerical techniques in solving partial differential equations, which may arise in holographic problems, including, for instance holographic lattices \cite{Horowitz:2012ky,Rozali:2012es,Donos:2014yya}. 

The methods outlined here are applicable to the equations which may have singular points at the boundaries and to the problems with arbitrary boundary conditions. We intentionally avoid discussing any theory of Applied Mathematics, lying behind these methods since this is better explained in the excellent standard courses \cite{trefethen2000spectral,boyd2001chebyshev,briggs2000multigrid}. Most of the content here is in a big part copied from these books. Anyway, we tried to be more explicit when dealing with the systems of equations, and implementation of the boundary conditions. The subjects which are usually omitted in the standard courses, since they are very straightforward, but proved to be quite confusing when one is dealing with them for the first time. 

\section{Linear equations with constant coefficients}

Consider a differential equation with constant coefficients (Internal Equation)
\begin{equation}
\label{equ:lin_equ}
\mbox{IE:} \qquad
A \, \p_z^2 f(z) + B \, \p_z f(z) + C \, f(z) = R
\end{equation}
with boundary conditions (top and bottom)
\begin{align}
\label{equ:lin_bc}
\mbox{BCb:}& \qquad B_b \, \p_z f(z_b) + C_b \, f(z_b) = R_b \\
\mbox{BCt:}& \qquad B_t \, \p_z f(z_t) + C_t \, f(z_t) = R_t
\end{align}

The essence of Finite Difference Derivative (FDD) method is to turn this linear differential equation problem on the interval $z \in [z_b,z_t]$ into a system of linear algebraic equations. One does it by introducing the \textbf{grid} in the coordinate domain consisting of $N$ points:
\begin{equation}
\vec{z} = \{z_1, z_2, \dots, z_N \} \in [z_b,z_t], \qquad z_1 = z_b, \quad z_N = z_t
\end{equation}
The simplest example would be
\begin{equation}
\label{equ:grid}
z_i = z_b + \frac{z_t - z_b}{N-1} (i-1), \qquad i = 1 \dots N 
\end{equation}
Note that the boundary points \textbf{are included} in the grid.
\begin{equation}
z_1 \equiv z_b, \qquad z_N \equiv z_t 
\end{equation}

The function ($f$) and its derivatives ($\p_z f, \p_z^2 f$) are naturally promoted to be the \textbf{$N$-component vectors}, representing the corresponding values at the grid points:
\begin{align}
\label{equ:vec_f}
f(z) & \rar  \vec{f} & f_i &\equiv f(z_i) \\
\p_z f(z) & \rar \overrightarrow{\p_z f} & (\p_z f)_i &\equiv \p_z f(z_i) \\
\p^2_z f(z) & \rar \overrightarrow{\p^2_z f} & (\p^2_z f)_i &\equiv \p^2_z f(z_i) 
\end{align}

\subsection{Differrentiation matrices}

At this stage the question arises: Given the values of the function on the grid, how can one evaluate the values of its derivatives on the same grid? Firstly, note that because differentiation is a linear operation, the vectors $\lvec{\p_z f}$ and $\lvec{f}$ are related by the linear transformation. Therefore the matrices exist, such that
\begin{equation}
\lvec{\p_z f} = \mathbb{D}_z \cdot \lvec{f}
\end{equation}
and similarly
\begin{equation}
\lvec{\p^2_z f} = \mathbb{D}_{zz} \cdot \lvec{f}
\end{equation}
The $N\times N$ matrices $\mathbb{D}_z$ and $\mathbb{D}_{zz}$ are called \textbf{differentiation matrices}.

What is the explicit form of the differentiation matrix? Consider for example the simplest, \textit{nearest neighbour} scheme for the first derivative matrix $\mathbb{D}_z$. The prescription for the finite difference derivative in this case is:
\begin{equation}
(\p_z f)_i  = \frac{f_{i+1} - f_{i-1}}{z_{i+1} - z_{i-1}} = \frac{f_{i+1} - f_{i-1}}{2 \Delta z}, \qquad i \neq \{1,N\}.
\end{equation}
Clearly, this formula is inapplicable at the boundaries. On these points one has to use \textbf{one-sided derivatives} instead:
\begin{align}
(\p_z f)_1 &= \frac{-3 f_1 + 4 f_2 - f_3}{2 \Delta z} \\
(\p_z f)_N &= \frac{f_N - 4 f_{N-1} + 3 f_{N-2}}{2 \Delta z}
\end{align}
Combining these expressions together we can write down the explicit form of the differentiation matrix:
\begin{equation}
\label{equ:diff_mat}
\mathbb{D}_z = \frac{1}{2 \Delta z}\begin{pmatrix}
-3 & 4 & -1 & \dots & 0 & 0 & 0\\
-1 & 0 & 1 & \dots & 0 & 0 & 0\\
0 & -1 & 0 & \dots & 0 & 0 & 0 \\
\vdots & \vdots & \ddots & \ddots & \ddots  & \vdots & \vdots \\
0 & 0 & 0 & \dots & 0 & 1 & 0 \\
0 & 0 & 0 & \dots & -1 & 0 & 1 \\
0 & 0 & 0 & \dots & 3 & -4 & 1 
\end{pmatrix}
\end{equation}
Similarly one constructs the matrix $\mathbb{D}_{zz}$. 

Clearly, the nearest neighbour approximation is not the only possible choice for discretizing the derivative. One can use two, three or more nearest points in order to approximate the derivative with better accuracy. This leads to a denser differentiation matrix. For the details of the various approximations and their accuracy we refer the reader to the excellent tutorial ``The Numerical Method of Lines'' of Wolfram Mathematica \cite{Mathematica10}. 

\subsection{Operator of the internal equations}

Using the differentiation matrices, one can recast the differential equation \eqref{equ:lin_equ} as a linear system
\begin{equation}
\left(A \, \mathbb{D}_{zz}  + B \, \mathbb{D}_z + C \, \mathbb{I} \right) \cdot \vec{f} - R \, \vec{1} = 0,
\end{equation}
where $\mathbb{I}$ is $N\times N$ identity matrix and $\vec{1}$ is $N$ vector of unities. We can introduce the \textbf{linear operator} 
\begin{equation}
\mathbb{O} \equiv  A \, \mathbb{D}_{zz}  + B \, \mathbb{D}_z + C \, \mathbb{I}.
\end{equation}
Then the vector of the equations \eqref{equ:lin_equ} on the grid reads
\begin{equation}
\label{equ:internal_system}
\lvec{IE} : \quad \mathbb{O} \cdot \vec{f} = R \, \vec{1}
 \end{equation}

\subsection{Boundary conditions}

It is important to note here that the linear system \eqref{equ:internal_system} \textbf{is not} equivalent to the boundary value problem \eqref{equ:lin_equ}, \eqref{equ:lin_bc}, since it does not include the boundary conditions yet. Indeed, $(\lvec{IE})_1$ and $(\lvec{IE})_N$ are the equations \eqref{equ:lin_equ} evaluated on the endpoints, which we should substitute with the appropriate discretized boundary conditions \eqref{equ:lin_bc}. Moreover, in practice it often happens that the internal equations \eqref{equ:lin_equ} are singular on the boundaries and evaluating them on the endpoints doesn't make any sense. 

In order to discretize the boundary conditions \eqref{equ:lin_bc} one can follow the same procedure as for the internal equations. The equations \eqref{equ:lin_bc} (for all grid points $z_i$) can be represented as
\begin{align}
\lvec{BCb} &= \mathbb{O}_b \cdot{f} - R_b \, \vec{1}, & \lvec{BCt} &= \mathbb{O}_t \cdot{f} - R_t \, \vec{1} \\
\mathbb{O}_b & \equiv  B_b \, \mathbb{D}_z + C_b \, \mathbb{I} & \mathbb{O}_t & \equiv  B_t \, \mathbb{D}_z + C_t \, \mathbb{I}
\end{align}
We do not actually need the boundary conditions at \textbf{all} points, since we should only keep $(\lvec{BCb})_1$ for the condition at $z_b$ and $(\lvec{BCt})_N$ for the condition at $z_t$. More precisely, we need
\begin{align}
\label{equ:bc_system}
(\lvec{BCb})_1 &= (\mathbb{O}_b)_{1 j} (\vec{f})^j - R_b (\vec{1})_1, \\
(\lvec{BCt})_N &= (\mathbb{O}_t)_{N j} (\vec{f})^j - R_t (\vec{1})_N,
\end{align}
where the summation over $j$ is assumed and only the \textbf{first line} of $\mathbb{O}_b$ and \textbf{last line} of $\mathbb{O}_t$ are used. 

\subsection{Operator of the full problem}

One can merge \eqref{equ:internal_system} and \eqref{equ:bc_system} in a single system of $N$ linear equations, which represent the full boundary value problem \eqref{equ:lin_equ}, \eqref{equ:lin_bc}:
\begin{equation}
\label{equ:equ_vec}
\lvec{\mathrm{BVP}} : \qquad \begin{cases}
(\lvec{BCb})_1 \\
(\lvec{IE})_2 \\
\dots \\
(\lvec{IE})_{N-1} \\
(\lvec{BCt})_N
\end{cases} 
=
\begin{cases}
(\mathbb{O}_b)_{1 j} (\vec{f})^j - R_b \, (\vec{1})_1 \\
(\mathbb{O})_{2 j} (\vec{f})^j - R \, (\vec{1})_2 \\
\dots \\
(\mathbb{O})_{N-1 \, j} (\vec{f})^j - R \, (\vec{1})_{N-1} \\
(\mathbb{O}_t)_{N j} (\vec{f})^j - R_t \, (\vec{1})_N
\end{cases}
\equiv
\tilde{\mathbb{O}} \cdot \vec{f} - \vec{\tilde{R}}
\end{equation}
Here we introduced the \textbf{operator of the boundary value problem} (BVP operator) $\tilde{\mathbb{O}}$, which coincides with $\mathbb{O}$ everywhere except the first and the last lines, which are substituted from $\mathbb{O}_b$ and $\mathbb{O}_t$, respectively. Similarly, the vector of the right hand side $\vec{\tilde{R}}$ coincides with $R \, \vec{1}$ everywhere except first and last entry, where the values $R_b$ and $R_t$ are substituted.  

In the end of the day, the BVP problem \eqref{equ:lin_equ}, \eqref{equ:lin_bc} can be recast in the matrix form
\begin{equation}
\label{equ:linear_BVP}
\tilde{\mathbb{O}} \cdot \vec{f} - \vec{\tilde{R}} = 0,
\end{equation}
and can be solved by direct inversion of the BVP operator
\begin{equation}
\vec{f} = \tilde{\mathbb{O}}^{-1} \cdot \vec{\tilde{R}}
\end{equation}

\subsection{Assignment}
Solve the equation
\begin{equation}
f''(z) + \pi^2 f(z) = 0
\end{equation}
in the domain $z \in [0,1]$ with the boundary conditions
\begin{equation}
f(0) = 1, \qquad f'(1) ={} 0
\end{equation}
\textit{Solution}
\begin{equation}
f(z) = \cos(\pi z) 
\end{equation}

\section{\label{sec:variable_coef}Linear equations with variable coefficients}
In the previous section we have already encountered the situation where the values of the coefficients $\vec{R}$ in \eqref{equ:linear_BVP} are different inside the grid and on the endpoints. In this section we generalize this feature to the arbitrary variable coefficients on the grid.

Consider a linear differential equation with variable coefficients
\begin{equation}
\label{equ:lin_var_equ}
\mbox{IE:} \qquad
A(z) \, \p_z^2 f(z) + B(z) \, \p_z f(z) + C(z) \, f(z) = R(z)
\end{equation}
with boundary conditions
\begin{align}
\label{equ:lin_var_bc}
\mbox{BCb:}& \qquad B_b \, \p_z f(z_b) + C_b \, f(z_b) = R_b \\
\mbox{BCt:}& \qquad B_t \, \p_z f(z_t) + C_t \, f(z_t) = R_t
\end{align}

\subsection{Vectors of coefficients}
Now not only the function and its derivatives will become vectors of values on the grid, but also the coefficients of the equation. Similarly to \eqref{equ:vec_f} we get:
\begin{equation}
\label{equ:vec_a}
X(z)  \rar  \vec{X}, \qquad  X_i \equiv X(z_i), \qquad X \in \{A,B,C,R\}
\end{equation}

Since the coefficients take different values, one effectively has to solve \textit{different algebraic equations at every grid point}. 
Using the same notation as \eqref{equ:equ_vec}, we write down the resulting system of equations equivalent to the boundary value problem.
\begin{equation}
\label{equ:equ_vec_var}
\lvec{\mathrm{BVP}} = \begin{cases}
(\lvec{BCb})_1 \\
(\lvec{IE})_2 \\
\dots \\
(\lvec{IE})_{N-1} \\
(\lvec{BCt})_N
\end{cases} 
=
\begin{cases}
(0 \, \mathbb{D}_{zz} + B_b \, \mathbb{D}_z + C_b \, \mathbb{I})_{1 j} (\vec{f})^j - R_b  \\
(A_2 \, \mathbb{D}_{zz}  + B_2 \, \mathbb{D}_z + C_2 \, \mathbb{I})_{2 j} (\vec{f})^j - R_2  \\
\dots \\
(A_{N-1} \, \mathbb{D}_{zz}  + B_{N-1} \, \mathbb{D}_z + C_{N-1} \, \mathbb{I})_{N-1 \, j} (\vec{f})^j - R_{N-1}  \\
(0 \, \mathbb{D}_{zz} + B_t \, \mathbb{D}_z + C_t \, \mathbb{I})_{N j} (\vec{f})^j - R_t 
\end{cases} 
\end{equation}
We can rewrite it as a matrix equation like \eqref{equ:linear_BVP}
\begin{equation}
\label{equ:linear_BVP_var}
\tilde{\mathbb{O}} \cdot \vec{f} - \vec{\tilde{R}} = 0
\end{equation}
by defining the linear operator matrix line-wise, i.e.
\begin{equation}
\tilde{\mathbb{O}}_{i j} \equiv \tilde{A}_i  (\mathbb{D}_{zz})_{i j} + \tilde{B}_i  (\mathbb{D}_z)_{i j} + \tilde{C}_i  (\mathbb{I})_{i j}, \qquad i,j = 1 \dots N.
\end{equation}
Or in the matrix form
\begin{equation}
\label{equ:matrix_O}
\tilde{\mathbb{O}} \equiv diag(\vec{\tilde{A}}) \cdot \mathbb{D}_{zz} + diag(\vec{\tilde{B}})  \cdot \mathbb{D}_z + diag(\vec{\tilde{C}})  \cdot \mathbb{I},
\end{equation}
where $diag(\vec{\tilde{X}})$ is a diagonal matrix with the entries of the vector $\vec{\tilde{X}}$ on the diagonal and the coefficient vectors are defined as
\begin{equation}
\label{equ:coeff_arrays}
\begin{cases}
\tilde{X}_1 \equiv X_b \\
\tilde{X}_i \equiv X_i, \quad i = 2\dots N-1 \\
\tilde{X}_N \equiv X_t
\end{cases}
\qquad X \in \{A,B,C,R\}
\end{equation}
and we define $A_t = A_b \equiv 0$, since the boundary conditions must be first order. 

Note that now all the information about the boundary value problem, \textbf{including boundary conditions} is encoded in the set of coefficient vectors $\vec{\tilde{X}}$.

Once the BVP operator \eqref{equ:matrix_O} is constructed, the linear problem \eqref{equ:linear_BVP_var} can be solved by direct inversion:
\begin{equation}
\vec{f} = \tilde{\mathbb{O}}^{-1} \cdot \vec{\tilde{R}}
\end{equation}

\subsection{Assignment}
Solve the equation
\begin{equation}
f''(z) - \pi f'(z) \cot(\pi z) = 0
\end{equation}
in the domain $z \in [0,1]$ with the boundary conditions
\begin{equation}
f(0) = 1, \qquad f'(1) = 0.
\end{equation}
Note that the equation is singular at the boundary $z=1$.
\\
\textit{Solution}
\begin{equation}
f(z) = \cos(\pi z) 
\end{equation}

\section{\label{Sec:Nonlinear}Nonlinear equations}
In the previous section we considered the equations with coefficients depending on the coordinate $z$. It is straightforward to generalize this treatment to the case when the coefficients are dependent on the function itself -- the nonlinear differential equations. The essential step to be made is to set up the iterative procedure for the nonlinear equation, which relies on solving the linearized system at every step.

\subsection{Iterative solution to nonlinear differential equation: Newton method}
Consider the nonlinear equation
\begin{equation}
\label{equ:nonlin_equ}
\mathbf{E}\left[\p_z^2 F(z), \p_z F(z), F(z) , z \right] = G(z),
\end{equation}
where $\mathbf{E}$ is an arbitrary function.
Assume $F_0$ is an exact solution to this equation and $F_n$ is a close enough approximation to it. More precisely
\begin{equation}
\label{equ:F_approx}
F_n = F_0 + f, \qquad ||f||\ll 1,
\end{equation}
with some chosen norm $||*||$.
Applying $\mathbf{E}$ to both sides leads to 
\begin{equation}
\mathbf{E}\left[F_n \right] = G + \frac{\delta \mathbf{E}\left[F_n \right]}{\delta \p_z^2 F} \p_z^2 f(z) + \frac{\delta \mathbf{E}\left[F_n \right]}{\delta \p_z F} \p_z f(z) + \frac{\delta \mathbf{E}\left[F_n \right]}{\delta F} f(z) + O(f^2).
\end{equation}
It can be recast in familiar form
\begin{equation}
\label{equ:linearized_equ}
\mathbf{A}[F_n,z] \p_z^2 f(z) + \mathbf{B}[F_n,z] \p_z f(z) + \mathbf{C}[F_n,z] f(z) = \mathbf{R}[F_n,z],
\end{equation}
with
\begin{equation}
\mathbf{R}[F_n,z] \equiv  \mathbf{E}[F_n,z] - G(z).
\end{equation}

Once the linearized equation \eqref{equ:linearized_equ} is solved, one obtains the next, better, approximation to the solution:
\begin{equation}
F_{n+1} = F_{n} - f
\end{equation}
and the procedure is reiterated up to the point when the nonlinear equation is satisfied to the desired accuracy $\delta$, i.e.
\begin{equation}
||\mathbf{R}[F_n,z] || = || \mathbf{E}[F_n,z] - G(z) || \ll \delta
\end{equation}
This iterative method is known as \textbf{Newton method for iterative solution} of the nonlinear equation.

\subsection{Variable coefficients}
The only difference between the linearized equation \eqref{equ:linearized_equ} and the linear equation \eqref{equ:lin_var_equ} is the fact that the coefficients are now dependent not only on the coordinate $z$, but also on the values of the background function $F_n$ and its derivatives on the grid. Therefore, the evaluation of the coefficient vectors \eqref{equ:coeff_arrays} breaks into two steps. 

Firstly, given the $n$-th approximation $F_n (z_i)$ on the grid one uses a chosen method to obtain the values of $\p_z F_n(z_i)$ and $\p_z^2 F_n(z_i)$. For instance by applying the differentiation matrices \eqref{equ:diff_mat}. 

Then one substitutes the vectors $\lvec{F}, \lvec{\p_z F}, \lvec{\p_z^2 F}$ and the grid coordinates into the functions $\mathbf{A},\mathbf{B},\mathbf{C}$ and $\mathbf{R}$ in order to obtain the vectors of coefficients \eqref{equ:vec_a}:
\begin{equation}
\mathbf{X}[F,z] \rar \vec{X}, \qquad X_i \equiv \mathbf{X}[(\p_z^2 F)_i,(\p_z F)_i, F_i, z_i], \qquad X=\{A,B,C,R\}
\end{equation}
Note that again, in order to construct the BVP operator, the boundary conditions should be substituted at the endpoints as in \eqref{equ:coeff_arrays}. The nonlinear boundary conditions can be linearized following exactly the same procedure as for the main equation. For every step in the iteration procedure, the BVP operator is constructed as
\begin{equation}
\label{equ:matrix_O_nonlin}
\tilde{\mathbb{O}}[F_n] \equiv diag(\vec{\tilde{A}}[F_n]) \cdot \mathbb{D}_{zz} + diag(\vec{\tilde{B}}[F_n])  \cdot \mathbb{D}_z + diag(\vec{\tilde{C}}[F_n])  \cdot \mathbb{I}.
\end{equation}
and the subsequent approximation is obtained as
\begin{equation}
\label{equ:nonlinear_iteration}
\vec{F}_{n+1} = \vec{F}_n - \tilde{\mathbb{O}}[F_n]^{-1} \cdot \vec{\tilde{R}}[F_n]
\end{equation}

In the Newton method, the vectors of coefficients are recalculated at every iteration step. The \textit{pseudo-Newton methods} exist, which take advantage of the fact that $F_{n+1}$ is generically close to $F_n$ and allow to use some of the coefficients evaluated at the previous steps.

\subsection{Assignment}
Solve the nonlinear equation
\begin{equation}
F''(z) - F(z)^2 = 2 + z^4
\end{equation}
in the domain $z \in [0,1]$ with the boundary conditions
\begin{equation}
F(0) = 0, \qquad F(1) = 1.
\end{equation}
\textit{Solution}
\begin{equation}
F(z) = z^2 
\end{equation}
 
\section{\label{sec:System}System of equations}
In the previous sections we were dealing with a single equation on a single function. How are the outlined procedures modified in case of a system of the coupled differential equations on several functions?

Consider the linear system of $K$ equations on the functions $f^k(z)$, $k=1\dots K$. It can be represented as a $K$-valued vector differential equation
\begin{equation}
\label{equ:system}
\mathcal{A} \cdot \begin{pmatrix} \p_z^2 f^1(z) \\ \vdots \\ \p_z^2 f^K(z) \end{pmatrix} 
+ \mathcal{B} \cdot \begin{pmatrix} \p_z f^1(z) \\ \vdots \\ \p_z f^K(z) \end{pmatrix} 
+ \mathcal{C} \cdot \begin{pmatrix}  f^1(z) \\ \vdots \\  f^K(z) \end{pmatrix} 
=  \begin{pmatrix}  R^1(z) \\ \vdots \\  R^K(z) \end{pmatrix},
\end{equation}
where $\mathcal{A},\mathcal{B},\mathcal{C}$ are $K\times K$-matrices of (coordinate dependent) coefficients and $R^k$ is a $K$-vector of right hand sides.

The boundary conditions consist of $K$ equations on every boundary, characterized similarly to \eqref{equ:lin_var_bc} by the matrices of constant coefficients $\mathcal{B}_b,\mathcal{C}_b$ and $\mathcal{B}_t,\mathcal{C}_t$.

\subsection{Flattened vectors}

When we discretize this system on a lattice with $N$ nodes, every function $f^k$ (and its derivatives) turns to $N$-vector. Therefore $(f^1(z), \dots,f^K(z))^T$ turns into $K$-vector of $N$-vectors. This is not a very handy object. Instead we will arrange the values of all the functions on the grid as a single $K\cdot N$ \textbf{flattened} vector. 
\begin{equation}
\label{equ:flattening}
\begin{pmatrix}
\left(f^1(z_1), \dots, f^1(z_N) \right)^T \\
\vdots \\
\left(f^K(z_1), \dots, f^K(z_N) \right)^T 
\end{pmatrix} 
\xrightarrow{flatten} 
\begin{pmatrix}
f^1(z_1) \\ \vdots \\ f^1(z_N) \\ f^2(z_1) \\ \vdots \\ f^K(z_N)
\end{pmatrix}
\end{equation}

\subsection{Enlarged differential matrix}

Given this data structure for the values of the functions, we wish to have similar objects for the derivatives. This requires defining the enlarged $K\cdot N \times K\cdot N$ differentiation matrices. Since the values of one function do not affect the derivatives of the other function, it is clear that the shape of the enlarged differentiation matrices will be block diagonal. For instance
\begin{equation}
\label{equ:diffmat_system}
\begin{pmatrix}
\p_z f^1(z_1) \\ \vdots \\ \p_z f^1(z_N) \\ \p_z f^2(z_1) \\ \vdots \\ \p_z f^K(z_N)
\end{pmatrix}
=
\begin{pmatrix}
\mathbb{D}_z & 0 & \cdots & 0 \\
0 & \mathbb{D}_z & \cdots & 0 \\
\vdots & \vdots & \ddots & \vdots \\
0 & 0 & \cdots & \mathbb{D}_z
\end{pmatrix}
\cdot
\begin{pmatrix}
f^1(z_1) \\ \vdots \\ f^1(z_N) \\ f^2(z_1) \\ \vdots \\ f^K(z_N)
\end{pmatrix},
\end{equation}
where $\mathbb{D}_z$ is the familiar $(N\times N)$ differentiation matrix on the grid \eqref{equ:diff_mat}. In a more concise notation, the enlarged differentiation matrix $\overline{\mathbb{D}}_z$ can be defined via the \textbf{Kronecker product}
\begin{equation}
\label{equ:kronecker_system}
\overline{\mathbb{D}}_z \equiv \mathbb{I}_{K\times K} \otimes {\mathbb{D}_z}
\end{equation}
Here and in what follows we will use bar in order to distinguish the objects in the space of flattened vectors. 

Note that Kronecker product is not commutative, there is a useful ``rule of thumb'' to remember the order of terms. It coincides with the order of indices in the ``vector of vectors'' structure in \eqref{equ:flattening} before flattening: in order to access the value $f^k(z_i)$ one has to first take the $k$-th line and then the $i$-th value in this line.

\subsection{Coefficients}
The situation with coefficient $(K\times K)$-matrices $\mathcal{A}, \dots$ is a bit more convoluted. Upon discretization on the grid every entry of the matrix turns into an $N$-vector of corresponding values at the lattice nodes. These vectors should multiply the rows of the corresponding differentiation matrices as in \eqref{equ:matrix_O}. Therefore we similarly turn them into the diagonal $(N\times N)$-matrices. Eventually, the coefficients in the equations get arranged in the $(K\times K)$-block matrices with diagonal $(N\times N)$ blocks. These block matrices will further multiply the enlarged differentiation matrices in order to provide the $(K\cdot N\times K\cdot N)$ BVP operator. 

Consider as example a system of 2 equations:
\begin{equation}
\label{equ:example_system}
\begin{matrix}
IE_1: \\ IE_2:
\end{matrix}
\quad
\begin{cases}
 A_{11} \p_z^2 f_1 + B_{11} \p_z f_1 + B_{12} \p_z f_2  = R_1 \\
 A_{22} \p_z^2 f_2 + B_{21} \p_z f_1   = R_2
\end{cases} 
\end{equation}
It can be recast in the form \eqref{equ:system} with the coefficient matrices
\begin{equation}
\mathcal{A} = 
\begin{pmatrix}
A_{11} & 0 \\
0 & A_{22} 
\end{pmatrix}, 
\qquad
\mathcal{B} = 
\begin{pmatrix}
B_{11} & B_{12} \\
B_{21} & 0 
\end{pmatrix}, 
\qquad
\mathcal{C} = 
\begin{pmatrix}
0 & 0 \\
0 & 0
\end{pmatrix}.{}
\end{equation}
Upon discretization, using the techniques developed in Sec.\ref{sec:variable_coef} we can represent the first equation in \eqref{equ:example_system} as an $N$-vector equation
\begin{equation}
\label{equ:example_old}
diag(\lvec{A_{11}}) \cdot \mathbb{D}_{zz} \cdot \vec{f_1} + diag(\lvec{B_{11}}) \cdot \mathbb{D}_z \cdot \vec{f_1} +  diag(\lvec{B_{12}}) \cdot \mathbb{D}_z \cdot \vec{f_2} = \lvec{R_1}
\end{equation}

Using the notation of the enlarged matrices we work with the whole system at once. The coefficient matrices turn into the block matrices
\begin{equation}
\label{equ:example_coefficients}
\overline{\mathcal{A}} = 
\begin{pmatrix}
diag(\lvec{A_{11}}) & 0 \\
0 & diag(\lvec{A_{22}}) 
\end{pmatrix}, 
\quad
\overline{\mathcal{B}} = 
\begin{pmatrix}
diag(\lvec{B_{11}}) & diag(\lvec{B_{12}}) \\
diag(\lvec{B_{21}}) & 0 
\end{pmatrix},
\quad
\overline{\mathcal{C}} = 
\begin{pmatrix}
0 & 0 \\
0 & 0 
\end{pmatrix}.
\end{equation}
And the differentiation matrices are
\begin{equation}
\overline{\mathbb{D}_z^2} = 
\begin{pmatrix}
\mathbb{D}_z^2 & 0 \\
0 & \mathbb{D}_z^2 
\end{pmatrix}, 
\qquad
\overline{\mathbb{D}_z} = 
\begin{pmatrix}
\mathbb{D}_z & 0 \\
0 & \mathbb{D}_z 
\end{pmatrix}, 
\qquad
\overline{\mathbb{I}} = 
\begin{pmatrix}
\mathbb{I} & 0 \\
0 & \mathbb{I} 
\end{pmatrix}.
\end{equation}
The full system of equations takes the form
\begin{equation}
\label{equ:example_new}
\overline{IE}: \qquad
\left(\overline{\mathcal{A}} \cdot \overline{\mathbb{D}_z^2} + \overline{\mathcal{B}} \cdot \overline{\mathbb{D}_z} + \overline{\mathcal{C}} \cdot \overline{\mathbb{I}} \right) \cdot \overline{f}  = \overline{R},
\end{equation}
where $\overline{f} = flatten[(\vec{f_1}, \vec{f_2})^T]$ is a flattened vector \eqref{equ:flattening} and $\overline{R} = flatten[(\lvec{R_1}, \lvec{R_2})^T]$. By straightforward multiplication of block matrices one can check that the first $N$ lines of \eqref{equ:example_new} do indeed coincide with \eqref{equ:example_old}. The advantage of the flattened notation is that the whole system is represented as a single matrix equation \eqref{equ:example_new}.

\subsection{Boundary conditions}
The question about implementation of the boundary conditions becomes more subtle as well. The system of $K$ differential equations on an interval requires $2K$ boundary conditions: one per function on two boundaries. 
In complete analogy with the treatment of Sec.\ref{sec:variable_coef}, in order to implement the boundary conditions as in \eqref{equ:coeff_arrays} one has to substitute the first and last elements in the coefficient $N$-vectors $\vec{X}_{\alpha\beta}$ in \eqref{equ:example_coefficients} with the corresponding values of $(X_b)_{\alpha\beta}, (X_t)_{\alpha\beta}$. This would produce the effective coefficient vectors $\vec{\tilde{X}}_{\alpha \beta}$, which can be used to construct the BVP operator and right hand side:
\begin{gather}
\label{equ:bvp_system}
\overline{BVP}: \qquad \left(\overline{\tilde{\mathcal{A}}} \cdot \overline{\mathbb{D}_z^2} + \overline{\tilde{\mathcal{B}}} \cdot \overline{\mathbb{D}_z} + \overline{\tilde{\mathcal{C}}} \cdot \overline{\mathbb{I}} \right) \cdot \overline{f}  = \overline{\tilde{R}}, \\
\overline{\tilde{\mathcal{X}}} = 
\begin{pmatrix}
diag(\vec{\tilde{X}}_{11}) & \dots & diag(\vec{\tilde{X}}_{1 K}) \\
\vdots & \ddots & \vdots \\
diag(\vec{\tilde{X}}_{K1}) & \dots & diag(\vec{\tilde{X}}_{K K})
\end{pmatrix}
, \qquad  \mathcal{X} \in \{\mathcal{A},\mathcal{B},\mathcal{C} \}
\end{gather}

It is instructive to figure out what does this operation mean for the flattened vector of equations \eqref{equ:example_new}. Since this flattened vector has exactly the same structure as $\overline{f}$ in \eqref{equ:flattening}, we can \textit{unflatten} it and represent as a $K$-vector of $N$-vectors of equations
\begin{equation}
\overline{IE} \equiv 
\begin{pmatrix}
IE^1(z_1) \\ \vdots \\ IE^1(z_N) \\ IE^2(z_1) \\ \vdots \\ IE^K(z_N)
\end{pmatrix} 
 \xrightarrow{\mathrm{unflatten}} 
\begin{pmatrix}
\left(IE^1(z_1), \dots, IE^1(z_N) \right)^T \\
\vdots \\
\left(IE^K(z_1), \dots, IE^K(z_N) \right)^T 
\end{pmatrix} 
\end{equation}
Then implementing the boundary conditions means that every entry $IE^\alpha(z_1)$ gets substituted with $BC_b^\alpha$ and $IE^\alpha(z_N)$ with $BC_t^\alpha$. It is most easily achieved by working on the \textit{second level} of the ``vector of vectors'' structure and flattening it in the end
\begin{equation}
\overline{BVP} \equiv 
\begin{pmatrix}
BC_b^1 \\ IE^1(z_2) \\ \vdots \\ BC_t^1 \\BC_b^2 \\ IE^2(z_2) \\ \vdots \\ BC_t^K
\end{pmatrix} 
  = \mathrm{flatten}
\begin{pmatrix}
\left(BC_b^1, IE^1(z_2), \dots, BC_t^1 \right)^T \\
\vdots \\
\left(BC_b^K,IE^K(z_2), \dots, BC_t^K \right)^T 
\end{pmatrix} 
\end{equation}

\section{Partial Differential Equations}
In many regards the implementation of the partial differential equations is similar to the case of a system of equations discussed above. The obvious additional complication is the variety of the distinct differentiation operators and the structure of the corresponding differentiation matrices. We will consider the case of 2-dimensional PDE, generalization to the higher dimensions is straightforward. In 2-dimensions (denote them $x$ and $y$) there are 6 distinct differentiation operators, including identity \footnote{In case of a single dimension, considered above, there were only 3 of them, in 3-dimensional equation there are 10, and so on.} 
\begin{equation}
\label{equ:2Ddiffs}
\p_p \in \{\p_{xx}, \p_{yy}, \p_{xy}, \p_{x}, \p_{y} , 1 \}, 
\end{equation}
 Therefore a 2D partial differential equation can be characterized by a set of 6 coefficients $X^p(x,y)$ (which replace $A,B$ and $C$ used in 1D \eqref{equ:lin_var_equ}) and a right hand side:
 \begin{equation}
\label{equ:PDE}
 PDE: \qquad \sum_p^6 X^p(x,y) \, \p_p f(x,y)  = R(x,y)
 \end{equation}
 
\subsection{Flattened vectors} 
Now we are dealing with a 2-dimensional grid:
\begin{align}
x &\rar x_i, \qquad i=1\dots N \\
y &\rar y_j, \qquad j=1 \dots M 
\end{align}
When discretized on this grid, the function becomes a level 2 array, which we can be expressed as a ($N\times M$) matrix
\begin{equation}
\label{equ:2D_discretization}
f(x,y) \rar \vec{\vec{f}}, \qquad  \vec{\vec{f}} \equiv \begin{pmatrix}
f(x_1,y_1) & \dots & f(x_1, y_M) \\
\vdots &  & \vdots \\
 f(x_N,y_1) & \dots & f(x_N, y_M) 
\end{pmatrix}, 
\qquad 
f_{ij} \equiv f(x_i, y_j)
\end{equation}
As in Sec.\ref{sec:System} we have to turn this data to a vector first, therefore we introduce the flattened $N\cdot M$-vector
\begin{equation}
\label{equ:2Dflatten}
\overline{f} \equiv \mathrm{flatten}(\vec{\vec{f}}) = 
\begin{pmatrix}
f(x_1,y_1) \\
\vdots \\
f(x_1,y_M) \\
f(x_2,y_1) \\
\vdots \\
f(x_N,y_M)
\end{pmatrix}
\end{equation}

\subsection{2D differentiation matrices}
Similarly to \eqref{equ:diffmat_system} we have to introduce the enlarged differentiation matrices, suitable for this data structure. It is easy to understand how the $y$-derivative should look like. Since it does not mix the functions at different $x$-coordinates, it acts as a block diagonal matrix on a flattened vector
\begin{equation}
\label{equ:diffmat_y_2D}
\begin{pmatrix}
\p_y f(x_1, y_1) \\ \vdots \\ \p_y f(x_1, y_M) \\ \p_y f(x_2, y_1) \\ \vdots \\ \p_y f(x_N, y_M)
\end{pmatrix}
=
\begin{pmatrix}
\mathbb{D}_y & 0 & \cdots & 0 \\
0 & \mathbb{D}_y & \cdots & 0 \\
\vdots & \vdots & \ddots & \vdots \\
0 & 0 & \cdots & \mathbb{D}_y
\end{pmatrix}
\cdot
\begin{pmatrix}
f(x_1, y_1) \\ \vdots \\ f(x_1, y_M) \\ f(x_2, y_1) \\ \vdots \\ f(x_N, y_M)
\end{pmatrix}.
\end{equation}
Therefore, in complete analogy with \eqref{equ:diffmat_system}, we can define it as a Kronecker product
\begin{equation}
\overline{\mathbb{D}_y} = \mathbb{I}_{N\times N} \otimes \mathbb{D}_y
\end{equation}

The $x$-derivative might seem less trivial. For instance, the $x$-derivatives at point $x_i$,  $ \lvec{(\p_x f)}_i \equiv \{(\p_x f)_{i1}, \dots, (\p_x f)_{i M}\}$, are obtained as a linear combination of $M$-vectors $\vec{f}_{i-1}$ and $\vec{f}_{i+1}$. Therefore the enlarged differentiation matrix should act on vectors as (compare to \eqref{equ:diff_mat})
\begin{equation}
\label{equ:diffmat_x_2D}
\begin{pmatrix}
\p_x f(x_1, y_1) \\ \vdots \\ \p_x f(x_1, y_M) \\ \p_x f(x_2, y_1) \\ \vdots \\ \p_x f(x_N, y_M)
\end{pmatrix}
=
\frac{1}{2 \Delta x} 
\begin{pmatrix}
-3 \mathbb{I} & 4 \mathbb{I} & - \mathbb{I} & \dots & 0 \\
- \mathbb{I} & 0 &  \mathbb{I} & \dots & 0 \\
0 & - \mathbb{I} & 0 & \dots & 0 \\
\vdots & \vdots & \ddots & \ddots& \vdots \\
0 & 0 & 0 & \dots &  \mathbb{I}
\end{pmatrix}
\cdot
\begin{pmatrix}
f(x_1, y_1) \\ \vdots \\ f(x_1, y_M) \\ f(x_2, y_1) \\ \vdots \\ f(x_N, y_M)
\end{pmatrix},
\end{equation}
Where $\mathbb{I}$ is a $M\times M$ identity matrix. Conveniently, this differentiation matrix can also be represented as a Kronecker product:
\begin{equation}
\overline{\mathbb{D}_x} = \mathbb{D}_x \otimes \mathbb{I}_{M\times M}
\end{equation}
Similarly, all the other operators in \eqref{equ:2Ddiffs} are represented by the matrices obtained as Kronecker products:
\begin{align}
\label{equ:kronecker_2D}
\overline{\mathbb{D}_{xx}} &= \mathbb{D}_{xx} \otimes \mathbb{I}_{M\times M}, & 
\overline{\mathbb{D}_{yy}} &= \mathbb{I}_{N\times N} \otimes \mathbb{D}_{yy}, & 
\overline{\mathbb{D}_{xy}} &= \mathbb{D}_{x} \otimes \mathbb{D}_{y},
\\
\overline{\mathbb{D}_{x}} &= \mathbb{D}_{x} \otimes \mathbb{I}_{M\times M},&
\overline{\mathbb{D}_{y}} &= \mathbb{I}_{N\times N} \otimes \mathbb{D}_{y},& 
\overline{\mathbb{I}} &=\mathbb{I}_{N\times N} \otimes \mathbb{I}_{M\times M}.
\end{align}
The ``rule of thumb'' is the same as in \eqref{equ:kronecker_system}. The first term in the product is an operator acting on the first index in the matrix form $\vec{\vec{f}}$ \eqref{equ:2Dflatten}, which corresponds to $x$-coordinate and the second -- operator acting on the second index, $y$. Analogously one can construct differentiation matrices in 3 dimensions. Theses will be the Kronecker products of 3 terms, corresponding to the operators acting on the 3 different coordinates.

\subsection{Boundary conditions}
Upon the discretization on the 2D grid the coefficients $X^p(x,y)$ turn to the matrices as in \eqref{equ:2D_discretization}. Similarly to Sec.\ref{sec:variable_coef} we implement the boundary conditions by substituting the corresponding entries in the matrices of coefficients.

In case of 2-dimensional problem one has 4 boundary conditions: top, bottom, left and right, characterized by the equations with coefficients $X^p_t(x), X^p_b(x), X^p_l(y)$ and $X^p_r(y)$ ($p=1\dots 6$), respectively. The corresponding boundaries on the grid are located at $y=y_M$ (top), $y=y_1$ (bottom), $x=x_1$ (left) and $x=x_N$ (right). The modified array of coefficients, which includes boundary conditions, is most easily constructed in the ``matrix'' notation:
\begin{gather}
\vec{\vec{X}} \rar \vec{\vec{\tilde{X}}} \\
\vec{\vec{X}} \equiv \begin{pmatrix}
X(x_1,y_1) & X(x_1,y_2) & \dots &  X(x_1, y_{M-1}) & X(x_1, y_M) \\
X(x_2,y_1) & X(x_2,y_2) & \dots &  X(x_2, y_{M-1}) & X(x_2, y_M) \\
\vdots &  \vdots& \ddots & \vdots & \vdots\\
X(x_{N-1},y_1) & X(x_{N-1},y_2) & \dots & X(x_{N-1}, y_{M-1})  & X(x_{N-1}, y_M) \\
X(x_{N},y_1) & X(x_{N},y_2) & \dots & X(x_{N}, y_{M-1})  & X(x_{N}, y_M) 
\end{pmatrix} \\
\vec{\vec{\tilde{X}}} \equiv \begin{pmatrix}
X_l(y_1) & X_l(y_2) & \dots & X_l(y_{M-1}) & X_l(y_M) \\
X_b(x_2) & X(x_2,y_2) & \dots & X(x_2, y_{M-1}) & X_t(x_2) \\
\vdots &  \vdots& \ddots & \vdots & \vdots\\
X_b(x_{N-1}) & X(x_{N-1},y_2) & \dots & X(x_{N-1}, y_{M-1})  & X_t(x_{N-1}) \\
X_r(y_1) & X_r(y_2) & \dots & X_r(y_{M-1})  & X_r(y_M) 
\end{pmatrix},
\end{gather}
As a result, the first and last rows and the first and last columns of the coefficient matrix $\vec{\vec{X}}$ are substituted by the corresponding boundary conditions in $\vec{\vec{\tilde{X}}}$. Same operation is performed with the right hand side terms $\vec{\vec{R}}$. It should be noted here that consistency of the boundary conditions requires in the corners:
\begin{equation}
X_l(y_1) = X_b(x_1), \quad X_r(y_1) = X_b(x_N), \quad X_r(y_M) = X_t(x_{N}), \quad X_l(y_M) = X_t(x_1)
\end{equation}

\subsection{BVP operator}

In the end of the day we need to construct the BVP operator. In complete analogy with Sec.\ref{sec:variable_coef} this is done by row-wise multiplication of the differentiation matrices with the coefficients \eqref{equ:matrix_O}. Some care should be taken to match the coefficients at given grid point with the derivatives in this point, but this is accounted for by the way we construct the differentiation matrices \eqref{equ:diffmat_y_2D}, \eqref{equ:diffmat_x_2D}, \eqref{equ:kronecker_2D}, which return the flattened vectors of the derivatives, compatible with the flattened vectors of coefficients 
\begin{equation}
\overline{\tilde{X}} \equiv \mathrm{flatten}\Big(\vec{\vec{\tilde{X}}}\Big).  
\end{equation}
Therefore the BVP operator for the partial differential equation \eqref{equ:PDE} with corresponding boundary conditions is computed as
\begin{equation}
\tilde{\mathbb{O}} = \sum_p^6 diag\Big(\overline{{\tilde{X}}^p}\Big) \cdot \overline{\mathbb{D}}_p
\end{equation}
And the boundary value problem can be solved by a direct inversion
\begin{equation}
\label{equ:inversion_2D}
\overline{f} = \tilde{\mathbb{O}}^{-1} \cdot \overline{{\tilde{R}}}
\end{equation}

\subsection{Assignment}
Solve the partial differential equation
\begin{equation}
\p_x^2 F(x,y) + \p_y^2 F(x,y) + \pi^2 F(x,y) =0 
\end{equation}
in the domain $(x,y) \in [0,1] \times [0,1]$ with the boundary conditions
\begin{equation}
F(x,0) = \sin(\pi x), \quad \p_y F(x,1) = 0, \quad F(0,y)=0, \quad F(1,y) = 0.
\end{equation}
\textit{Solution}
\begin{equation}
F(x,y) = \sin(\pi x)
\end{equation}

\section{Periodic boundary conditions}
So far we have only addressed the boundary conditions which can be expressed as local equations on the functions and their derivatives at the edges of the calculation domain. The other important type of the boundary conditions are the periodic ones. The essence of the periodic boundaries is the \textbf{identification} of the endpoints of the calculation interval. Once such an identification is performed, no extra information, like special equations for the boundaries, is needed. 

Take the homogeneous calculation grid on the interval $z \in [z_b,z_t)$ with identified endpoints. Similarly to \eqref{equ:grid} we can define
\begin{equation}
\label{equ:periodic_grid}
z_i = z_b + \frac{z_t - z_b}{N'} (i-1), \qquad i = 1 \dots N'. 
\end{equation}
Note that now the boundary point $z_t$ \textbf{is not included} in the grid. Reason for this is due to the fact that we identify the function at the endpoints $z_t$ and $z_b$ 
\begin{equation}
f(z_t) \equiv f(z_b),
\end{equation}
So there is no reason to solve for the value $f(z_b)$ twice, while looking for a solution. Note also that the new definition \eqref{equ:periodic_grid} leads to $N'$ points in the calculation grid (excluding $z_t$).

\subsection{Differentiation matrices}

The identification $f(z_t)$ and $f(z_b)$ requires a new type of the differentiation matrices. Consider the example with nearest neighbour approximation used in \eqref{equ:diff_mat}: $(\p_z f)_i = (f_{i+1} - f_{i-1})/2 \Delta z$. Inside the calculation domain no modifications are needed, since for $i$ in $2\dots N'-1$ the neighbouring points $f_{i-1}$ and $f_{i+1}$ exist. What will happen near the boundaries? When $i=1$ we would have $(\p_z f)_1 = (f_{2} - f_{1-1})/2 \Delta z$. There is no point $z_0$ in the domain, so $f(z_0)$ doesn't make sense. Instead, keeping in mind that we identified $f_1$ with $f_{N'+1}$, we substitute $f_{1-1}$ with $f_{N'}$ and get
\begin{equation}
(\p_z f)_1 = \frac{(f_{2} - f_{N'})}{2 \Delta z} \qquad (\p_z f)_{N'} = \frac{(f_{1} - f_{N'-1})}{2 \Delta z}.
\end{equation}
Same ``cyclic rule'' is applied for any differences with any number of neighbours. The differentiation matrix corresponding to this rule looks simple
\begin{equation}
\label{equ:periodic_diff_mat}
\mathbb{D}_z = \frac{1}{2 \Delta z}\begin{pmatrix}
0 & 1 & 0 & \dots & 0 & 0 & -1\\
-1 & 0 & 1 & \dots & 0 & 0 & 0\\
0 & -1 & 0 & \dots & 0 & 0 & 0 \\
\vdots & \vdots & \ddots & \ddots & \ddots  & \vdots & \vdots \\
0 & 0 & 0 & \dots & 0 & 1 & 0 \\
0 & 0 & 0 & \dots & -1 & 0 & 1 \\
1 & 0 & 0 & \dots & 0 & -1 & 0 
\end{pmatrix}
\end{equation}
Note the values in the upper right and lower left corners. 

Conveniently, once the periodic differentiation matrices are found, the implementation of the periodic boundary conditions is completed. Since the boundaries of the periodic domain are not in any way different from any internal points, one has to solve the same equations of motion there. Therefore there is no need to substitute endpoints of the coefficient vectors as it was done in \eqref{equ:coeff_arrays}. In the case of multiple dimensions, the appropriate matrices are chosen for each coordinate and then the enlarged matrices are again created as a Kronecker product.

\subsection{Assignment}
Solve the partial differential equation
\begin{equation}
\p_x^2 F(x,y) + \p_y^2 F(x,y) + 4\pi^2 F(x,y) =0 
\end{equation}
in the domain $(x,y) \in [0,1] \times [0,1]$ with the boundary conditions
\begin{equation}
F(x,0) = \sin(2 \pi x), \quad \p_y F(x,1) = 0, \quad \mbox{periodic in $x$}.
\end{equation}
\textit{Solution}
\begin{equation}
F(x,y) = \sin(2\pi x)
\end{equation}

\section{\label{sec:Relaxation} Relaxation and preconditioning}

All the preceding sections explained how to reduce the different kinds of the boundary value problems on the discrete grid to the linear matrix equations \eqref{equ:linear_BVP}, \eqref{equ:linear_BVP_var}, \eqref{equ:bvp_system}, \eqref{equ:inversion_2D}, or the iterative series of them in case of nonlinear BVP \eqref{equ:nonlinear_iteration}. The most obvious way to solve such a system is to directly invert the linear operator. Unfortunately in practice this can be extremely demanding, since the matrices to be inverted can become large for large multidimensional grids  and, for the higher order discretization schemes, dense. In this case the iterative relaxation approach can be useful.

\subsection{Relaxation}

Consider the linear system
\begin{equation}
\label{equ:linear_sys}
\mathbb{O} \cdot \vec{f} = \vec{G}.
\end{equation}
The essence of the relaxation method is to substitute \eqref{equ:linear_sys} with the time evolution equation
\begin{equation}
\label{equ:linear_time_evol}
\p_t \vec{f} = \mathbb{O} \cdot \vec{f} - \vec{G}.
\end{equation}
Clearly, the time evolution stops when \eqref{equ:linear_sys} is satisfied. It's instructive to study in more detail though, how exactly the solution is approached.

Assume that $f^0$ is an exact $t$-independent solution $\mathbb{O} \cdot \vec{f^0} \equiv \vec{G}, \p_t \vec{f^0} = 0$ at the later stage of the evolution \eqref{equ:linear_time_evol} we can represent the time-dependent solution as $\vec{f}(t)= \vec{f^0} + \vec{\delta f} (t)$, where time-dependent residue $\vec{\delta f}(t)$ is small. Plugging this back into \eqref{equ:linear_time_evol} we find
\begin{equation}
\label{equ:time_evolution_delta}
\p_t \, \vec{\delta f} = \mathbb{O} \cdot \vec{\delta f}.
\end{equation}

Consider the eigenfunctions $g^k$ of the operator $\mathbb{O}$ with the eigenvalues $\lambda^k$:
\begin{equation}
\mathbb{O}\cdot g^k = \lambda^k g^k.
\end{equation}
One can expand the residual function $\delta f$ in a basis of these eigenfunctions 
\begin{equation}
\delta f(t) = \sum_k c^k(t) g^k
\end{equation}
Plugging this in \eqref{equ:time_evolution_delta} we get simply
\begin{equation}
\p_t c^k(t) = \lambda^k c^k \qquad \Rightarrow \qquad c^k(t) \sim \exp(\lambda^k t).
\end{equation}
It is remarkable, that for an elliptic (i.e. Laplace) equation with a negatively definite operator with $\lambda^k < 0, \forall k$ the components of the residual function decay exponentially, therefore the relaxation it this case does actually converge to the static solution! Notice that the speed of convergence is set by the \textbf{lowest} eigenvalue $\delta f \sim \exp(\lambda_{\mathrm{min}})$, $\lambda_{\mathrm{min}} = - \mathrm{Min}[|\lambda^k|, \forall k]$

In practice one discretizes the time derivative in \eqref{equ:linear_time_evol} and setups the iterative procedure:

\begin{equation}
\label{equ:relax_iteration}
 f^{n+1}  =  f^n + \delta t \, \mathbb{O} \cdot f^n - \delta t \vec{G}.
\end{equation}
The main advantage of the relaxation method becomes obvious here: one does not need to invert $\mathbb{O}$ matrix at any point of the calculation!

It can be shown \cite{boyd2001chebyshev} that in order for the iteration to be numerically stable the value of the time step $\delta t$ should be limited by the \textbf{highest} eigenvalue $\lambda_{\mathrm{max}}$ of $\mathbb{O}$
\begin{equation}
\delta t \sim \frac{1}{\lambda_{\mathrm{max}}},
\end{equation}
Therefore the number of iterations needed to achieve the solution, set by the slowest mode is proportional to $\lambda_{\mathrm{max}}/\lambda_{\mathrm{min}}$. This is a serious drawback since for $N\times N$ matrix $\mathbb{O}$ this ratio is typically of order $N^2$. Therefore the trivial relaxation procedure \eqref{equ:relax_iteration} is extremely ineffective.

\subsection{Preconditioning}

The relaxation procedure would work much faster if one would be able to use the operator with $\lambda_{\mathrm{max}}/\lambda_{\mathrm{min}} \approx 1$. One way to improve the situation is to note that instead of the relaxation equation \eqref{equ:linear_time_evol} one can equally well use
\begin{equation}
\label{equ:linear_time_evol_precond}
\p_t \vec{f} = - \hat{\mathbb{O}}^{-1} \cdot \left(\mathbb{O} \cdot \vec{f} - \vec{G} \right),
\end{equation}
where $\hat{\mathbb{O}}$ is an arbitrary linear operator. Indeed, similar to \eqref{equ:linear_time_evol}, the evolution \eqref{equ:linear_time_evol_precond} will only stop when the solution to \eqref{equ:linear_sys} is achieved. The iterative procedure will now take the form
\begin{equation}
\label{equ:relax_iteration_prec}
 f^{n+1}  =  f^n - \delta t \,  \hat{\mathbb{O}}^{-1} \cdot \mathbb{O} \cdot f^n + \delta t  \hat{\mathbb{O}}^{-1} \cdot \vec{G},
\end{equation}
and the time-step will be limited now by the highest eigenvalue of the regulated operator $\hat{\mathbb{O}}^{-1} \cdot \mathbb{O}$. This is the essence of the \textbf{preconditioning} procedure and the matrix $\hat{\mathbb{O}}$ is a \textbf{preconditioner matrix}. 

It is easy to figure out that the perfect preconditioner, which will set $\lambda_{\mathrm{max}}/\lambda_{\mathrm{min}} = 1$ is an operator itself: $\hat{\mathbb{O}} = \mathbb{O}$. In this case the time step can be chosen as large as $\delta t = 1$ and the iteration \eqref{equ:relax_iteration_prec} degenerates to a direct solution of \eqref{equ:linear_sys} in one step. This is again not extremely practical, since inversion of the operator $\mathbb{O}$ is exactly something that we'd like to avoid.

But it's clear now how to use preconditioning to one's advantage. The preconditioner should be chosen in such a way that the highest eigenvalue of the product $\hat{\mathbb{O}}^{-1} \cdot \mathbb{O}$ is as close to unity as possible, while simultaneously the matrix $\hat{\mathbb{O}}$ is easily invertable. In case when $\mathbb{O}$ is constructed using some high order finite difference approximation, the good choice is to use the BVP operator of the same system, but constructed using \textbf{low} order finite difference approximation, for instance the nearest neighbour. This is known as an Orszag preconditioning \cite{boyd2001chebyshev}. This preconditioner approximates well the highest eigenvalues of $\mathbb{O}$, leading to $\delta t \approx 1$, but is also easily invertable being a sparse matrix due to the nearest neighbour finite difference scheme. The practical advantage of the Orszag preconditioning is also in the fact that since the operator $\hat{\mathbb{O}}$ has the same coefficients as $\mathbb{O}$, the effects of the singular terms in the equations gets diminished in the combination $\hat{\mathbb{O}}^{-1} \cdot \mathbb{O}$. Importantly, the low accuracy of the derivative approximation in $\hat{\mathbb{O}}$ doesn't affect the accuracy of the final solution, which is dictated by $\mathbb{O}$.

Importantly, the general analysis of the convergence of the relaxation schemes \cite{boyd2001chebyshev} shows that in any case the time step should be less then 1. For completeness, referring the reader to \cite{boyd2001chebyshev} for the details, we should mention here that one should use
\begin{equation}
\tau \leq \frac{4}{7}
\end{equation}
for a the stable iteration.

The practical recipe therefore is to use some high order derivative discretization scheme for the operator $\mathbb{O}$ in order to achieve high accuracy of the solution, but use the low order scheme for $\hat{\mathbb{O}}$, making it easily invertable. Then, just a few step of the iteration \eqref{equ:relax_iteration_prec} are enough to get the solution with desired precision.
  
Another practical advantage of the preconditioned relaxation outlined above shows up when one deals with the nonlinear equations discussed in Sec.\ref{Sec:Nonlinear}. Reason is that the solution of a nonlinear equation requires iteration anyway and there is no point to solve the equation exactly at every step. Therefore the nonlinear iteration can be easily combined with the relaxation iteration. Form this point of view one can regard the relaxation with preconditioning as a special version of a pseudo-Newton method, where at every step one inverts a certain approximation to $\mathbb{O}$ instead of the operator itself.

\subsection{Assignment}
Solve the problem of from one of the previous Sections with order 6 approximation for the derivatives in the equation operator. Use the direct inversion, then trivial relaxation \eqref{equ:relax_iteration} and in the end the relaxation with Orszag preconditioning. Compare the efficiencies of the applied methods.

\section{Pseudospectral method}

As it has been pointed out in the previous Section, the accuracy of the final solution is governed by the accuracy of the derivatives approximation used in the operator $\mathbb{O}$ in the right hand side of \eqref{equ:linear_time_evol}. It is therefore important to develop a procedure to approximate the derivatives accurately.

\subsection{Pseudospectral collocation}

The \textbf{pseudospectral collocation} method resides on the fact that instead of representing the function via its values on the grid, one can represent it as a series in the basis functions on a particular interval, giving the required values on the grid points. Consider for instance the interval $\theta \in [0, 2 \pi)$ with periodic boundary conditions. In these conditions the function can be represented as a Fourier series
\begin{equation}
\label{equ:fourier_series}
f(\theta) = \sum_k \hat{v}_k e^{i k \theta}
\end{equation}
Given the values of the function on a homogeneous grid $v_i = f(\theta_i)$ one can unambiguously evaluate the coefficients $\hat{v}_k$. Once the coefficients are known, the derivative of the series \eqref{equ:fourier_series} can be evaluated exactly:
\begin{equation}
w_i \equiv \p_\theta f(\theta_i) = \sum_k i k \hat{v}_k e^{i k \theta_i}
\end{equation}
Given that in order to figure out the coefficients $c_n$ the values of function on the whole grid are used, the differentiation matrices corresponding to the pseudospectral collocation method are dense. The method might be seen as N-th order finite difference derivative for $N$-point grid.

Similar expansion can be done for the interval with the boundaries $z\in[-1,1]$. in this case the basis functions are Chebyshev polynomials $T_k(z)$ and the unknown function can be represented as $F(z) = \sum V_k T_k$. The differential matrices are similarly dense in this case.

It looks like the pseudospectral methods deliver better accuracy for the price of having very inconvenient form of the differentiation matrices. This can be tolerated if one uses the relaxation discussed in Sec.\ref{sec:Relaxation}, since in this case there is no need to invert the matrix. But the bigger advantage of pseudospectral approach is unveiled once the efficient Fourier transform technique is used instead of the differentiation matrix multiplication.

\subsection{Pseudospectral collocation and Fourier transform}

Indeed, on a circle the coefficients of the Fourier series \eqref{equ:fourier_series} can be evaluated using the discrete Fourier transform. Given the values of the function on the homogeneous grid $v_i$, the Fourier coefficients are \cite{trefethen2000spectral}:
\begin{equation}
\label{equ:fourier}
\hat{v}_k= \delta \theta \sum_{j=1}^N e^{-i k \theta_j} v_j, k = -\frac{N}{2}+1, \dots, \frac{N}{2}
\end{equation}
and the Fourier coefficients for the derivative are simply
\begin{equation}
\hat{w}_k = i k \hat{v}_k.
\end{equation}
In order to obtain the values of the derivative on the grid points one has to make another, inverse Fourier transform
\begin{equation}
\label{equ:inv_fourier}
w_j = \frac{1}{2 \pi} \sum_{k=-N/2 +1}^{N/2} e^{i k \theta_j} \hat{w}_k, \qquad j = 1,\dots, N.
\end{equation}
The great advantage of the pseudospectral method is the fact that one can perform the operations \eqref{equ:fourier} and \eqref{equ:inv_fourier} efficiently using Fast Fourier Transform (FFT) algorithm and avoid using the nasty differentiation matrices all together. 

Similarly, one can use FFT in order to evaluate the derivative of the function on an interval $z \in [-1,1]$ represented as a series of Chebyshev polynomials. The distinctive feature of the Chebyshev polynomials on an interval is the fact that they can be directly related to the the harmonic functions on a circle. Given one identifies $z \equiv \cos(\theta)$ the n-th order polynomial is represented as
\begin{equation}
T_n (z) = \cos(n \theta), \qquad z \equiv \cos(\theta)
\end{equation}
\begin{figure}[ht]
\center
\includegraphics[width=0.4 \linewidth]{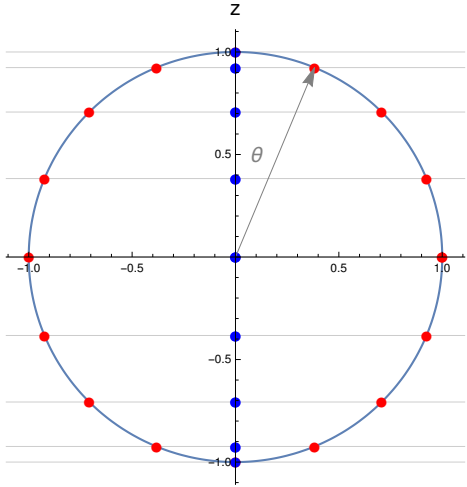}
\caption{\label{Fig:Cheb}Relation between homogeneous grid on $\theta \in [0,2 \pi)$ and Chebyshev grid on $z \in [-1,1]$}
\end{figure}

Therefore one can setup a one-to-one correspondence between the values of a symmetric function on a homogeneous grid on a circle
\begin{equation}
v_i \equiv f(\theta) = \sum \hat{v}_k \cos(k \theta_i), \qquad \theta_i =0, \delta \theta, 2 \delta \theta, \dots 
\end{equation}
and the values of a function on a ``Chebishev grid'' (see Fig.\ref{Fig:Cheb}) on an interval
\begin{equation}
V_i \equiv F(z) = \sum \hat{V}_k T_k(z_i), \qquad z_i =1, \cos(\delta \theta), \cos(2 \delta \theta), \dots 
\end{equation},
with the same set of coefficients $\hat{V}_k = \hat{v}_k$.

After this identification is done one can evaluate the derivatives of $f(\theta)$ with the FFT, and once it is done, relate it to the derivatives of $F(z)$:
\begin{equation}
\p_z F(z) = \p_z f(\theta) = \frac{\p z}{\p \theta} \p_\theta f(\theta) = \frac{\p_\theta f(\theta)}{\sqrt{1-z^2}}.{}
\end{equation}
This formula will work for all the point on the interval excluding endpoint $z=1, z=-1$, where the derivative is obtained using l'H\^opital's rule.

This is the final ingredient for a pseudospectral method. In the end of the day we see, that  it is possible to setup the very efficient procedure using the relaxation technique, low order difference preconditioner and high order pseudospectral approximation to the BVP operator $\mathbb{O}$, implemented through the FFTs. 

\section{Conclusion and implementation}
The outlined techniques one can reduce the nonlinear differential equation boundary value problem to the problem in linear algebra. In this form is it relatively straightforward to implement these procedures in one's favorite computing software. The relaxation procedure and the pseudospectral collocation technique require no more then the efficient sparse linear solver and fast Fourier transform. These routines can be find in any contemporary computation package or library including Mathematica, MATLAB, Python, FORTRAN etc. 
We do not discuss the implementation here leaving the choice to the reader. 
The methods discussed above, implemented in Wolfram Mathematica \cite{Mathematica10}, were successfully applied to several numerical projects in applied AdS/CFT including 1-dimensional and 2-dimensional problems \cite{Mott,Discom,Andrade:2017leb,Andrade:2015iyf,Krikun:2015tga,Andrade:2017cnc,Gorsky:2015pra} and proved to be effective.

\acknowledgments
I appreciate contribution and enthusiastic support of the participants of the Numerical Study group: Aurelio Romero-Bermudez, Philippe Sabella-Garnier, Floris Balm and Koenraad Schalm. I'm also grateful to Tomas Andrade in collaboration with whom most of the numerical projects were completed, where I handled the methods discussed here.

\bibliographystyle{JHEP-2}
\bibliography{inhom_stripes_lattice}

\end{document}